\documentclass[draft]{agujournal2019}
\usepackage{url} 
\usepackage{lineno}
\usepackage[inline]{trackchanges} 
\usepackage{soul}

\usepackage{graphicx}
\usepackage{amsmath}
\usepackage{amsfonts}
\usepackage{mathtools}

\usepackage{ragged2e}
\justifying

\draftfalse

\journalname{JGR: Space Physics}

\begin{document}

%
%


\title{Bayesian inference of quasi-linear radial diffusion parameters using Van Allen Probes}

%
%




\authors{Rakesh Sarma\affil{1}, Mandar Chandorkar\affil{1}, Irina Zhelavskaya\affil{2,3}, Yuri Shprits\affil{2,3}, Alexander Drozdov\affil{4}, Enrico Camporeale\affil{1,5,6}}


\affiliation{1}{Centrum Wiskunde \& Informatica, Amsterdam, The Netherlands}
\affiliation{2}{GFZ German Research Centre For Geosciences, Potsdam, Germany}
\affiliation{3}{Institute of Physics and Astronomy, University of Potsdam, Germany}
\affiliation{4}{University of California, Los Angeles, CA, USA}
\affiliation{5}{CIRES, University of Colorado, Boulder, CO,USA}
\affiliation{6}{NOAA, Space Weather Prediction Center, Boulder, CO, USA}




\correspondingauthor{Rakesh Sarma}{rakesh@cwi.nl}




\begin{keypoints}
\item We present the first application of Bayesian parameter estimation to the problem of quasi-linear diffusion in the radiation belts
\item The Bayesian approach allows the problem to be cast in probabilistic terms and for ensemble simulations to be run
\item An improved accuracy is demonstrated when compared against standard deterministic models
\end{keypoints}

%
%

%
%


\begin{abstract}
The Van Allen radiation belts in the magnetosphere have been extensively studied using models based on radial diffusion theory, which is based on a quasi-linear approach with prescribed inner and outer boundary conditions. The 1-d diffusion model requires the knowledge of a diffusion coefficient and an electron loss timescale, which are typically parameterized in terms of various quantities such as the spatial ($L$) coordinate or a geomagnetic index (for example, $Kp$). These terms are empirically derived, not directly measurable, and hence are not known precisely, due to the inherent non-linearity of the process and the variable boundary conditions. In this work, we demonstrate a probabilistic approach by inferring the values of the diffusion and loss term parameters, along with their uncertainty, in a Bayesian framework, where identification is obtained using the Van Allen Probe measurements. Our results show that the probabilistic approach statistically improves the performance of the model, compared to the parameterization employed in the literature. 
\end{abstract}

%
%

%


%
%
%
%

\section{Introduction}

The Van Allen radiation belts consist of energetic electrons and protons originating from the solar wind, which are trapped by the Earth's magnetic field.
The dynamics of these particles are affected by an interplay of different mechanisms, including local acceleration and losses due to wave-particle interactions and external injections \cite{Shprits2008a, Shprits2008b, reeves2013,ukhorskiy2012}. Understanding and forecasting the radiation belt particles is an essential part of Space Weather, since these particles can interfere with the satellites orbiting around the Earth and cause data loss. Hence, the estimation of the flux distribution in the space-time domain is a long-standing challenging problem in the space physics community \cite{Reeves2003, Shprits2008a, Shprits2008b}. 

The standard way of studying radiation belt particles dynamics is through a quasi-linear approach, that dates back to the seminal paper of \citeA{kennel1966} and has been routinely applied to magnetospheric particle since the 1970's \cite{schulz1969,lanzerotti1970,schulz2012}.  

The quasi-linear approach studies the particles' dynamics based on their quasi-periodic orbits with respect to the field lines of the Earth's magnetic field. Specifically, the cyclotron motion (or gyration) is the circular motion of particles around the field line, the bounce motion is due to the mirror force and along the field lines from one hemisphere to the other, and the drift motion is the orbit in the east or westward direction. These orbits are associated to the conservation of adiabatic invariants, respectively named the first ($\mu$), second ($K$) and third ($L^*$) invariant \cite{Roederer1970}. The evolution of the particle density is then studied as a diffusive process in the adiabatic invariant space, where the effect of resonant wave-particle interactions is described through diffusion coefficients. The third adiabatic invariant is associated with the longest timescale, and it can be violated by wave-particle interactions with Ultra-Low Frequency (ULF) waves. Therefore, as a first approximation, the system can be studied as radially diffusive, assuming the other two invariants to be conserved. In this study, we estimate the phase space density (PSD) based on the case of pure radial diffusion, which has been extensively investigated in the literature.  

The radial diffusion formulation involves a parametric representation of the diffusion coefficient and the electron loss timescale, which are generally formulated as varying in $L$ (the spatial coordinate) and $Kp$ (a geomagnetic index used as a proxy for the amplitude of geomagnetic perturbations). 

The most extensively employed diffusion rate parameterization is obtained by \citeA{Brautigam2000} based on the October, 1990 storm. There have been other developments where the radial diffusion coefficient is evaluated using various approaches. \citeA{Fei2006} constructed time dependent diffusion terms from MHD simulations of September 1998 storm using ULF electric and magnetic field power spectral density. In \citeA{Ozeke2012, Ozeke2014}, the diffusion coefficient is expressed as a sum of two terms due to azimuthal electric and compressional magnetic fields and expressions for these two terms are derived. A data-driven approach is employed to determine the radial diffusion coefficients in \citeA{Su2015}. The importance of the electron loss timescale in the radial diffusion modeling during storm-time is demonstrated in \citeA{Shprits2004}, where the interplay between inward radial diffusion and loss terms on the PSD is investigated. \citeA{Summers2007} obtained the loss timescale due to the combined effects of chorus, plasmaspheric hiss, and EMIC waves. \citeA{Shprits2007} has parameterized the loss timescale due to chorus waves as a function of the geomagnetic index, $AE$. Further extension to include the effect of multiple storms is investigated in \citeA{Tu2009} with an internal heating term. In \citeA{Ali2016}, the magnetic and electric field measurements from the Van Allen Probes are used to compute power spectral densities of both components, which are used with the \citeA{Fei2006} formulation to obtain magnetic and electric components of the radial diffusion components. 

It is important to emphasize that even in three-dimensional studies of the radiation belt that solve the diffusion equation in the whole adiabatic invariant space, hence taking into account energy and pitch-angle scattering, the radial diffusion coefficient is still often parameterized using the \citeauthor{Brautigam2000} formula (see, e.g., \citeA{Subbotin2009,su2010,tu2013,bourdarie2012,welling2012verification})

However, a single parameterization might not generalise well for different geomagnetic conditions. 
Moreover, the diffusion problem is significantly influenced by the boundary conditions, hence a deterministic parameterization might not be adequate to represent the uncertainties due to variable particle injections at the boundary. 
The aim of the current investigation is to obtain a probabilistic representation of the PSD by introducing uncertainties in the parametric representation of coefficients in the underlying partial differential equation.
This is one of the first applications of the Bayesian framework approach to parameter estimation of the radial diffusion equation. In the literature, data assimilation with an Extended Kalman filter technique is employed for determining the lifetime of electrons by \citeA{Kondrashov2007}. 

In this study, we perform Bayesian parameter identification approach for all the terms defining the 1-d diffusion equation. Once the diffusion coefficient and the electron time loss are defined as probability density functions, one can run an ensemble of simulations by sampling different values of the parameters and hence being able to estimate the uncertainty of the output PSD \cite{camporeale2016, camporeale2019challenge}.  
We use a data-driven representation of the input parameters, by employing Van Allen Probes data to identify the parameter distribution in a Bayesian setting \cite{spence2013}.  

The manuscript is structured in the following parts: Section 2 provides an introduction to the radial diffusion model and the chosen parameterization of coefficients. In Section 3, the framework for uncertainty propagation and the Bayesian identification is discussed. The results, discussion and comparison to the Van Allen measurements and other parameterization in literature are discussed in Section 4. Finally, the findings and future research directions are discussed in Section 5. 

\section{Modeling of Radial Diffusion dynamics}
Several radiation belt models \cite{Li2004,Subbotin2009,Reeves2012, Albert2009} have been developed to quantify the radial transport. The trapped particles are quantified for given adiabatic invariants ($\mu,K,L$) at time $t$ with the PSD, $f(\mu,K,L,t)$. Under the assumption that the invariants ($\mu,K$) are conserved, the radial diffusion model is a one-dimensional model based on the modified Fokker-Planck equation \cite{Walt1970}, and is given by:
\begin{equation}
    \frac{\partial f}{\partial t}=L^2\frac{\partial}{\partial L}\bigg[D_{LL}L^{-2}\frac{\partial f}{\partial L}\bigg]-\frac{f}{\tau},
    \label{eq:Diffusion_1-D}
\end{equation}
where, $D_{LL}$ is the diffusion coefficient and $\tau$ is the loss timescale, which is essentially a correction term for unaccounted dynamics (such as pitch-angle and energy scattering). Various parameterizations for $D_{LL}$ and $\tau$ are available in literature. In particular, in this study, we refer to the seminal work on the statistical diffusion rate of $D_{LL}$ proposed by \citeA{Brautigam2000}, which models the geomagnetic storm of October 9 1990, given by:
\begin{equation}
    D_{LL}(L,t)= \alpha_D L^{\beta_D} 10^{b_D Kp(t)},
    \label{eq:DLLparam}
\end{equation}
where they use the values
$\alpha_D=4.73\times 10^{-10}$, $\beta_D=10.0$ and $b_D=0.506$.
\citeA{Brautigam2000} use a simple estimation of the diffusion coefficients but their parameterization works surprisingly well even nowadays. However, the equation that is presented here is only one part of the \citeA{Brautigam2000} radial diffusion coefficient, which is the so called \textit{electromagnetic} part. Use of additional \textit{electrostatic} part leads to incorrect simulation results as was shown by \citeA{Kim2011}. The choice of the $L$ and $Kp$-dependent parameterization for the diffusion coefficient is motivated by the practise in literature. In the later part of this manuscript, we show that this choice leads to close agreement of PSD with respect to Van Allen Probes measurements. It has also been reported in earlier investigations \cite{Ali2016} that $D_{LL}$ exhibited a weak energy dependence in the range of $\mu$ between 500 MeV/G and 5000 MeV/G. Moreover, in this study we seek to demonstrate the application of Bayesian calibration to the diffusion problem, which can be applied to other choices of $D_{LL}$ parameterization in future investigations.

For the electron lifetime $\tau$, we employ $Kp$ and $L-$dependent parameterization based on \citeA{Gu2012} (without energy dependence) inside the plasmasphere, while a model based on \citeA{Ozeke2014} is used outside the plasmasphere. The plasmapause position $L_{pp}$ is estimated using a recently developed Plasma density in the Inner magnetosphere Neural network-based Empirical (PINE) model \cite{Zhelavskaya2017}. The PINE density model is developed using neural networks and is trained on the electron density data set from the Van Allen Probes Electric and Magnetic Field Instrument Suite and Integrated Science (EMFISIS) \cite{Kletzing2013}. The model reconstructs the plasmasphere dynamics well (with a cross-correlation of 0.95 on the test set), and its global reconstructions of plasma density are in good agreement with the IMAGE EUV images of global distribution of He+. The MLT-averaged plasmapause position is calculated using the output of the PINE model by applying a density threshold of $40$ cm$^{-3}$ to separate the plasmasphere from the outside of the plasmasphere. The $Kp$ index is obtained from the OMNIWeb database. The parameterization for $\tau$ that we employ in \eqref{eq:Diffusion_1-D} is given by:  
\begin{align}
\begin{split}
    \tau(L,t) &= (\alpha_\tau + \beta_\tau L + b_\tau L^2)/Kp(t) \textrm{ for } L\leq L_{pp} \\
    &= c_\tau/Kp(t) \hspace{2.6cm} \textrm{ for } L>L_{pp}.
    \label{eq:tau_param}
\end{split}
\end{align}
We choose a 1-year period from October 2012 to September 2013 for the purpose of analysis. The initial and outer boundary conditions are interpolated from the Van Allen Probes data. As mentioned, \eqref{eq:Diffusion_1-D} is obtained for a constant value of $(\mu, K)$. In this study, we compare all the results with Van Allen measurements for $\mu = 700$ MeV/G and $K=0.0019$ $\textrm{G}^{0.5}\cdot$Re. In order to estimate the accuracy of our predictions, we use the relative error as a metric, given by:
\begin{equation}
    \epsilon = \frac{|f_{va}-f|}{f_{va}},
    \label{eq:rel_err}
\end{equation}
where $f_{va}$ is the PSD value obtained from the Van Allen Probes and $f$ is the PSD-estimate from \eqref{eq:Diffusion_1-D}. The absolute value of the discrepancy is used here, since we are interested in estimating the overall performance of the solver throughout the domain, which will be obtained by integrating this error across the domain. Now, as defined in Section 1, we are interested in obtaining an informed estimate on the parameters defining the coefficients $D_{LL}$ and $\tau$ in \eqref{eq:Diffusion_1-D}. In the next section, we introduce the Bayesian framework which is used to identify these parameters. 

\section{Bayesian framework for identification}
The Bayesian approach to the calibration of computer models was introduced in \citeA{kennedy2001}, where the term calibration refers to adjusting the free parameters in order for the model output to fit the observations. In our case, the forward model is represented by \eqref{eq:Diffusion_1-D}, that is solved numerically with a standard finite-difference scheme on a uniform grid in $(L,t)$. Because solving this equation numerically is relatively fast, we opt for a standard Markov-Chain Monte Carlo (MCMC, \citeA{brooks2011}) procedure to explore the space of unknown free parameters  in \eqref{eq:DLLparam} and \eqref{eq:tau_param}, that are collected in a multi-dimensional vector $\mathbf{\Lambda}$ defined as:
\begin{equation*}
    \mathbf{\Lambda} := (\alpha_D, \beta_D, b_D, \alpha_\tau, \beta_\tau, b_\tau, c_\tau).
\end{equation*}
In other words, $\mathbf{\Lambda}$ is the set of uncertain parameters that are to be identified from this investigation.
The ground truth for the PSD is taken from Van Allen Probes measurements, and we derive a Bayesian model trained over a data-set of 30 days from 01-Oct-2012 to 30-Oct-2012. The objective is to demonstrate that the method is generalised for time-periods on which it is not trained, hence the model will be tested for the rest of the year (November 2012 to September 2013). 
The vector of parameters $\mathbf{\Lambda}$ is treated as a random variable, meaning that it is associated to an (unknown) probability density. The scope of the Bayesian inference is to estimate the probability of $\mathbf{\Lambda}$, given the PSD observations, that we denote with $\mathbf{f}^+$. Hence, we can use the classical Bayes' rule  \cite{Gelman2004}:
\begin{equation}
    \mathbb{P}(\mathbf{\Lambda}\mid \mathbf{f}^{+}) \propto \mathbb{P}(\mathbf{f}^{+}\mid\mathbf{\Lambda})\cdot \mathbb{P}_0(\mathbf{\Lambda}).
    \label{eq:bayes}
\end{equation}
where, $\mathbb{P}(\mathbf{f}^{+}\mid\mathbf{\Lambda})$ is the \textit{likelihood} that defines the discrepancy between the model estimate and the Van Allen Probes measurements, for a given realization of $\mathbf{\Lambda}$. The term $\mathbb{P}_0(\mathbf{\Lambda})$ is the \textit{prior} distribution of $\mathbf{\Lambda}$, which encodes all prior physical information one might have about the parameters.
 $\mathbb{P}(\mathbf{\Lambda}\mid \mathbf{f}^{+})$ is called the \textit{posterior} distribution, which in general cannot be expressed in closed form, but can be sampled through a Monte Carlo procedure (see below). Finally, once a sufficient number of samples from the posterior distribution has been collected, an ensemble simulation can be obtained by propagating each realization of $\mathbf{\Lambda}$ through \eqref{eq:Diffusion_1-D} to obtain the posterior predictive distribution of the PSD.

\subsection{Prior, likelihood, and posterior}

\noindent As a first step, the priors on the parameter set $\mathbf{\Lambda}$ are defined. In the present investigation, we assume uniform priors, given by:
\begin{equation}
    \mathbf{\Lambda} \sim \mathcal{U}(\mathbf{\Lambda}_{min},\mathbf{\Lambda}_{max}) 
    \label{eq:prior}
\end{equation}
where, $\mathbf{\Lambda}_{min}$ and $\mathbf{\Lambda}_{max}$ are chosen such that a wide domain is defined. The bound for each parameter of $\mathbf{\Lambda}$ is shown in Table \ref{tab:UQ_bounds}. Existing parameter values identified in literature \cite{Brautigam2000, Ozeke2014} are used as a reference, such that parameterizations obtained in these investigations belong to the set of the defined bounds. 
\begin{table}[]
    \centering
    \caption{Upper and lower bounds of prior defined in \eqref{eq:prior} for seven parameters in $\mathbf{\Lambda}$}
    \begin{tabular}{c|c|c}
    Parameter & Lower bound & Upper bound  \\
    $\alpha_D$ & $0.0$ & $1.4\times 10^{-7}$ \\
    $\beta_D$ & $1.0$ & $45.0$ \\
    $b_D$ & $0.0$ & $13.0$ \\
    $\alpha_\tau/Kp$ & $0.0/Kp$ & $26.0/Kp$ \\
    $\beta_\tau$ & $0.0$ & $5.0$ \\
    $b_\tau$ & $0.0$ & $5.0$ \\
    $c_\tau$ & $0.0$ & $30.0$
    \end{tabular}
    \label{tab:UQ_bounds}
\end{table}

We employ a Gaussian likelihood, by introducing an additive random variable $\eta$, specified as an unbiased normal distribution with a covariance matrix $\Sigma$. Under the assumption that there are no modeling errors, the statistical model can then be written as:
\begin{align}
\begin{split}
    \mathbf{f}^+ &= \mathbf{f}(\mathbf{\Lambda})+\eta\\
    \eta &\sim \mathcal{N}(0,\Sigma),
    \label{eq:stat_model}
\end{split}
\end{align}
where $\mathbf{f}(\mathbf{\Lambda})$ is the output of the forward model \eqref{eq:Diffusion_1-D}, which gives the likelihood as:
\begin{equation}
    \mathbb{P}(\mathbf{f}^{+}\mid\mathbf{\Lambda}):=\mathbb{P}_\eta(\mathbf{f}^+-\mathbf{f}(\mathbf{\Lambda}))=\mid\Sigma\mid^{-n/2}\exp\bigg[-\frac{1}{2}\big(\mathbf{f}^+-\mathbf{f}(\mathbf{\Lambda})\big)^{T}\Sigma^{-1}\big(\mathbf{f}^+-\mathbf{f}(\mathbf{\Lambda})\big)\bigg],
    \label{eq:likelihood}
\end{equation}
where $|\Sigma|$ is the determinant of the covariance matrix and $n$ is the number of Van Allen probe measurements. 

For sampling the posterior, we use the Metropolis-Hastings Markov Chain Monte Carlo (MCMC) algorithm  \cite{Hastings1970}, with a reversible Markov random-walk. The algorithm performs a random walk by sampling from a proposal distribution with a prescribed variance. Each new sample distribution is conditional only on the current sample, hence the generated sequence of samples resembles a Markov chain. Once a sample is chosen, a ratio of the densities for the two consecutive samples is computed, which informs the iterator if the jump results in increase/decrease of the posterior density. If the jump results in a  probability higher than an acceptance limit, the sample is accepted with that probability or it is rejected. The proposals are selected from a normal distribution, with tuned variance and a total of 20000 samples, where initial 1000 samples discarded as \textit{burn-in} period, and a thinning factor of 2 \cite{Gelman2004} is applied for the remaining samples to improve independence. It is to be noted here that choice for the burn-in period and thinning factor is problem-specific and it is decided based on the convergence of the parameters. The standard checks of convergence and auto-correlation for MCMC were applied \cite{Hastings1970, Chib1995}.
\section{Results and discussion}
In this section, we show the application of the Bayesian inference for identifying the parameters in \eqref{eq:Diffusion_1-D}, which are updated with the Van Allen Probes measurements. The 7-parameter uncertain space given by $\mathbf{\Lambda}$ is propagated with the forward solver and then the Bayesian framework is applied. 
\subsection{Posterior distribution of diffusion and lifetime parameters}
\begin{figure}[h]
  \centering
  \begin{tabular}{@{}c@{}}
    \includegraphics[width=0.8\linewidth]{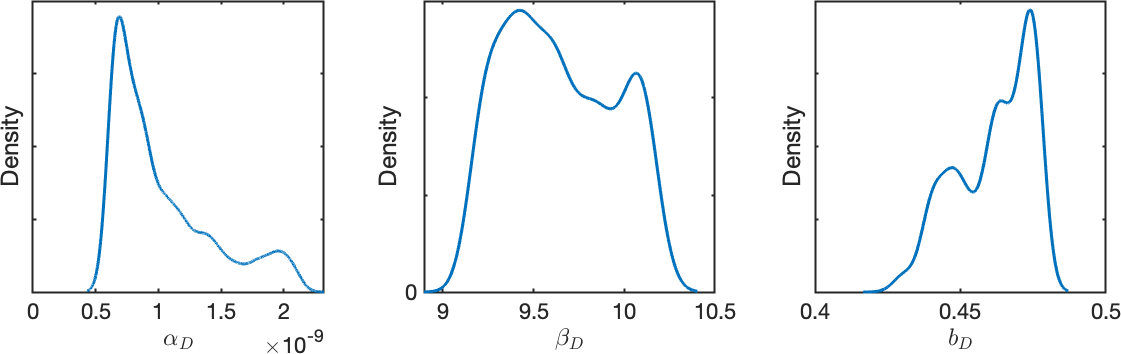} \\[\abovecaptionskip]
  \end{tabular}
  \hspace{0.5cm}
  \begin{tabular}{@{}c@{}}
    \includegraphics[width=0.8\linewidth]{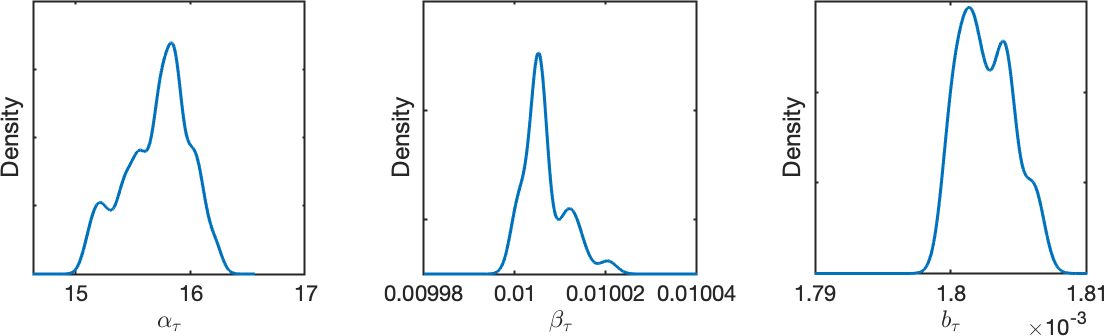} \\[\abovecaptionskip]
   \end{tabular}
  \hspace{0.5cm}
  \begin{tabular}{@{}c@{}}
    \includegraphics[width=0.235\linewidth]{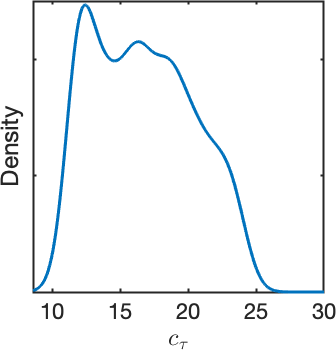} \\[\abovecaptionskip]
   \end{tabular}
  \caption{Posterior distributions of identified parameters $\mathbf{\Lambda}$ of $D_{LL}$ and $\tau$.}\label{fig:posteriors}
\end{figure}

Figure \ref{fig:posteriors} shows the posterior distributions of the identified parameters $\mathbf{\Lambda}$ of $D_{LL}$ and $\tau$. As mentioned, uniform priors are specified with wide domains, where for example, the prior of $\beta_D \sim \mathcal{U}(1,45)$. It is observed that all the identified posterior distributions have low variance, except for the parameter $c_\tau$, whose uncertainty is still high after identification, which shows that the data is not informative enough to infer $c_\tau$. It is also interesting to note that $c_\tau$ in \eqref{eq:tau_param} is employed above the plasmapause location $L_{pp}$, which implies that this term is mostly employed at higher values of $L$, where the flux from the boundary dominates the density predictions. As a result, the inferred values of $c_\tau$ strongly depend on the influx of particles at the outer boundary, which is highly uncertain and a deterministic representation is not possible.        

Furthermore, parameters $\beta_\tau$ and $b_\tau$ have values close to zero, which are coefficients for $L$ and $L^2$-dependence terms of $\tau$ in \eqref{eq:tau_param} respectively. This shows the variation of $\tau$ with respect to $L$ is minimal. It should also be mentioned here that other investigations of higher order variation of $\tau$ with $L$ also showed negligible dependence.    

\begin{figure}[h]
    \centering
    \includegraphics[width = 1\textwidth]{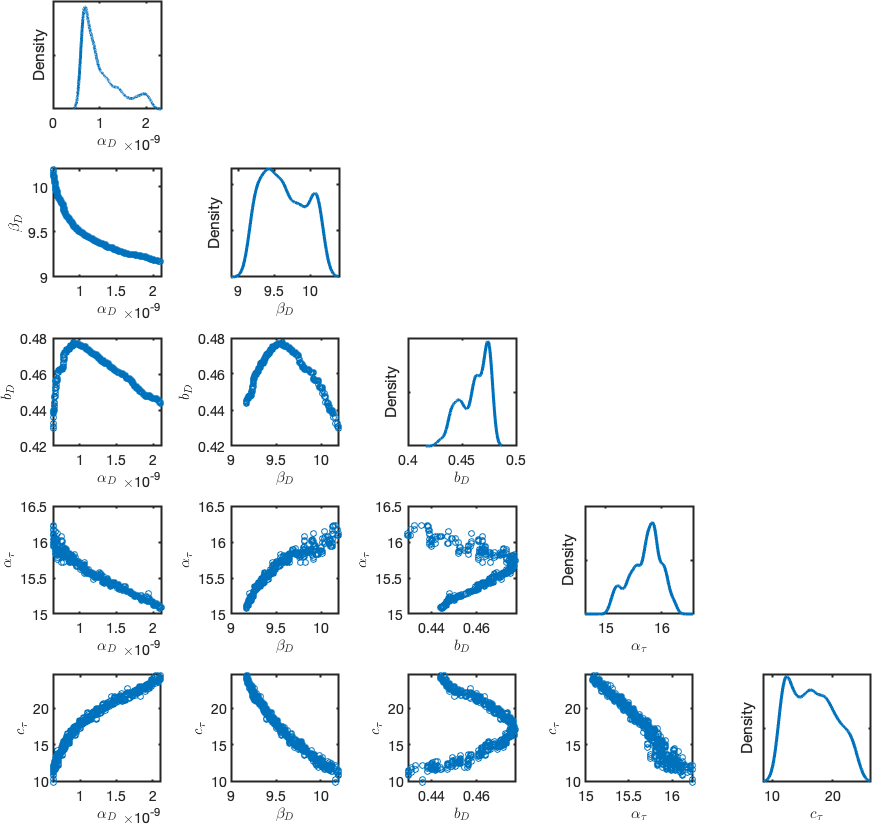}
    \caption{Identified 1-d posterior densities and 2-d marginal distributions for five parameters defining the diffusion and loss terms in \eqref{eq:Diffusion_1-D}.}
    \label{fig:posterior_marginal}
\end{figure}

In Figure \ref{fig:posterior_marginal}, 2-d marginal distributions of each of the parameter-pairs for five of the identified parameters in $\mathbf{\Lambda}$ are plotted along with the 1-d probability densities. The parameters $\beta_\tau$ and $b_\tau$ are not plotted here since their identified samples are close to zero. It is interesting to observe that clear correlations are identified between many parameter-pairs. Parameters $\alpha_D$ and $\beta_D$ have a clear negative correlation with an asymptotic trend. Furthermore, parameter-pairs $\alpha_D$ and $\alpha_\tau$, $\beta_D$ and $c_\tau$, $\alpha_\tau$ and $c_\tau$ have negative correlation, while $\alpha_D$ and $c_\tau$ seem to have a positive correlation. Other parameter pairs do not seem to show a clearly visible correlation behaviour.   

For the identified posteriors, we determine the maximum a posteriori (MAP) estimate of the probability distribution, which is the parameter-value corresponding to the point of maximum probability. Table \ref{tab:compare_MAP_Drozdov} compares the MAP estimate of the posteriors to the parameterization employed in the 1-d model of \citeA{Drozdov2017}, where two parameterizations for the 1-d diffusion problem are presented. In this study, we compare our estimates to the 1-d model based on \citeA{Brautigam2000} diffusion parameterization in \citeA{Drozdov2017}. Analytical $Kp$-dependent model based on \citeA{Shprits2005} was employed outside the plasmasphere, where $\tau = 3/Kp$ days is used. In order to take into account the effect of the variable outer boundary, \citeA{Shprits2006} used $\tau = 5/Kp$ as some of the loss was provided by the outward diffusion and longer lifetimes resulted in a better comparison with observations. Since we have a variable boundary in this investigation, we compare our estimates to a parameterization with a longer lifetime similar to \cite{Drozdov2017, Ozeke2014}, given by $\tau = 6/Kp$ days. It is to be noted that for the subsequent sections of this manuscript, we use the same choice of parameterization as reference. 

\begin{figure}[h]
    \centering
    \includegraphics[width = 1\textwidth]{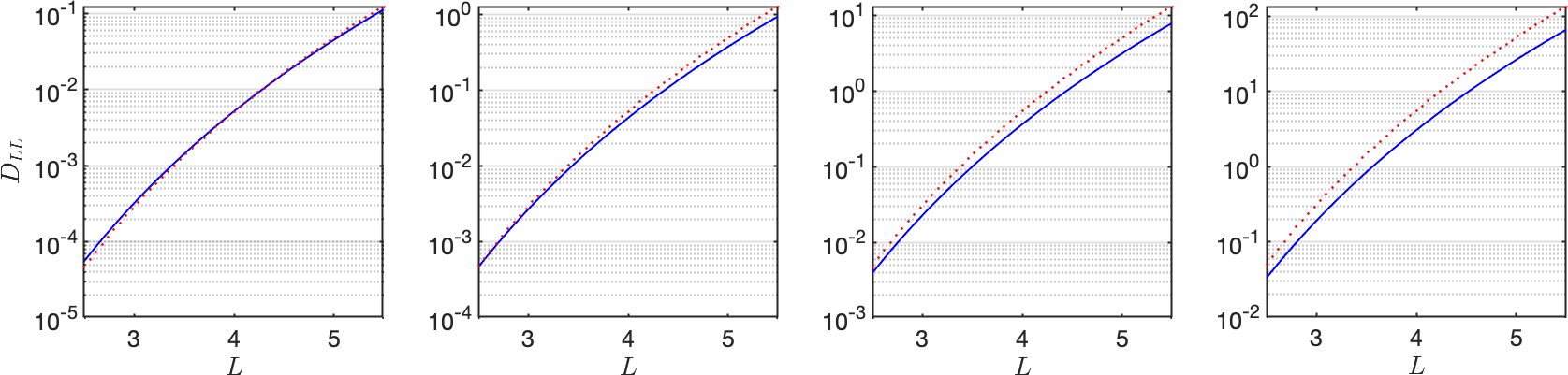}
    \caption{Probabilistic representation of $D_{LL}$ (in log scale) with the identified posteriors, with $Kp=2$, $4$, $6$ and $8$ from left to right. The red dots represent $D_{LL}$ obtained with \citeA{Brautigam2000} parameterization, while the blue solid lines represent the mean of the probabilistic $D_{LL}$ obtained from the posteriors.}
    \label{fig:DLL_probabilistic}
\end{figure}

Figure \ref{fig:DLL_probabilistic} compares the probabilistic $D_{LL}$ obtained with the identified posteriors shown in Figure \ref{fig:posteriors} to the $D_{LL}$ employed in \citeA{Brautigam2000}. At low values of $Kp$, the two predictions are close to each other, while the discrepancy between them progressively increases as $Kp$ is increased (note the vertical logarithmic scale). With \citeA{Brautigam2000} parameterization, $D_{LL}$ has higher values at higher $L$ with increase in $Kp$ compared to the probabilistic treatment. This could be related to the $Kp$-dependent $\tau$ parameterization, both above and below the plasmapause location. 

It is observed that the MAP estimates for the parameters of $D_{LL}$ namely $\alpha_D$, $\beta_D$ and $b_D$, have values close to the reference parameterizations. For $\beta_\tau$ and $b_\tau$, reference values do not exist and they both are close to zero. In \eqref{eq:tau_param}, $\alpha_\tau$ is $Kp$-dependent, while in the reference, it has a constant value of $10.0$ below and $6.0/Kp$ above the plasmapause location. With the current choice of parameterization, this term would be significantly smaller for high $Kp$ while being comparable to the reference value for low $Kp$. Furthermore, the term $c_\tau$ also significantly varies with respect to the reference value. As discussed already, $c_\tau$ has high variance, and the MAP estimate is higher than the reference. Hence it is observed that in terms of the MAP comparison, the lifetime estimates are giving longer time scales than that estimated in literature. This is clearly due to omission of local acceleration in the 1-d diffusion model. Similarly, the 1-d model in \citeA{Drozdov2017} underestimated the observations. Hence, the Bayesian parameter estimation is trying to compensate for the missing physical processes.  
\begin{table}[]
    \centering
    \caption{Comparison of parameterization based on the MAP estimate of the identified posterior to 1-d model in \citeA{Drozdov2017}}
    \begin{tabular}{c|c|c}
    Parameter & MAP estimate & Reference  \\
    $\alpha_D$ & $7.06\times 10^{-10}$ & $4.73\times 10^{-10}$ \\
    $\beta_D$ & $9.43$ & $10.0$ \\
    $b_D$ & $0.47$ & $0.51$ \\
    $\alpha_\tau/Kp$ & $15.84/Kp$ & $10.0$ \\ 
    $\beta_\tau$ & $0.01$ & - \\
    $b_\tau$ & $0.0018$ & - \\
    $c_\tau$ & $12.22$ & $6.0$ 
    \end{tabular}
    \label{tab:compare_MAP_Drozdov}
\end{table}
\subsection{Probabilistic representation of Phase Space Density}
Here we investigate the effect of the predicted uncertainties of the diffusion parameters on the evolution of the PSD. In practice, because the posterior distributions are derived by $\sim 10000$ collected samples, one can solve the diffusion equation \eqref{eq:Diffusion_1-D} for every realization of the parameters, hence generating an ensemble of runs from which to derive the uncertainty of the PSD. 
\subsubsection*{Probabilistic variation of density}
\begin{figure}[h]
  \centering
  \begin{tabular}{@{}c@{}}
    \includegraphics[width=1\linewidth]{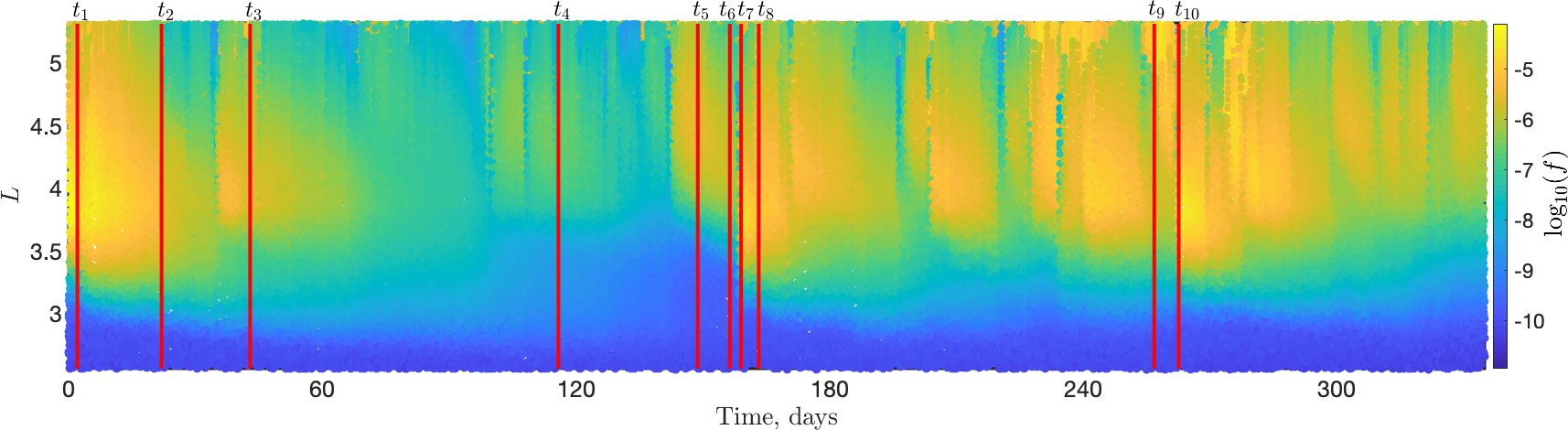} \\[\abovecaptionskip]
  \end{tabular} 
  \begin{tabular}{@{}c@{}}
    \includegraphics[width=0.97\linewidth]{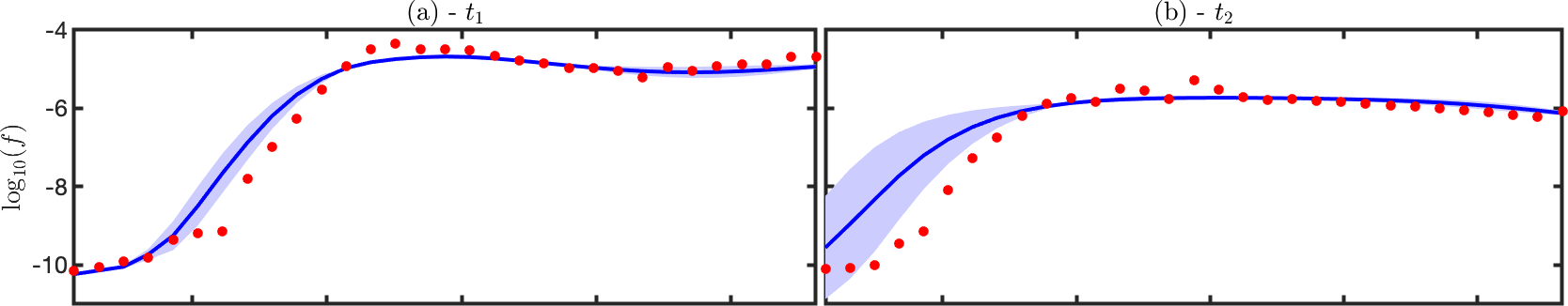} \\[\abovecaptionskip]
  \end{tabular}
  \begin{tabular}{@{}c@{}}
    \includegraphics[width=0.97\linewidth]{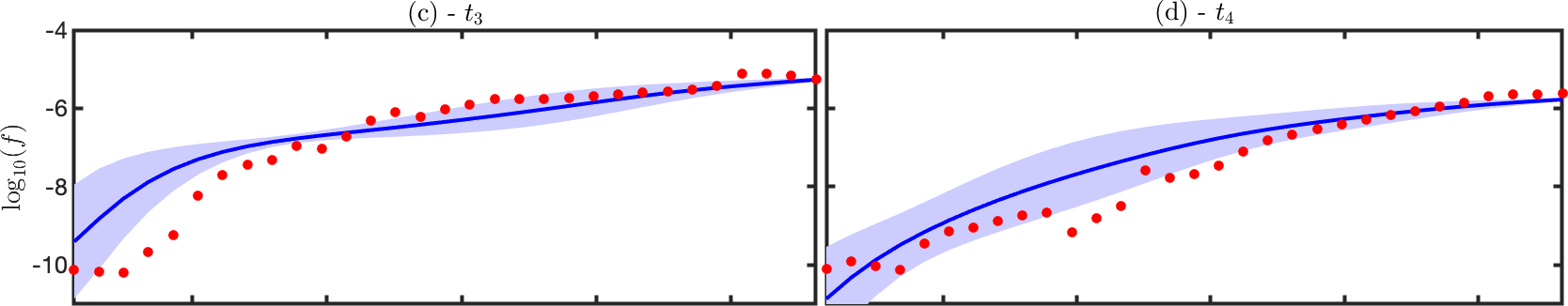} \\[\abovecaptionskip]
  \end{tabular}  
  \begin{tabular}{@{}c@{}}
    \includegraphics[width=0.97\linewidth]{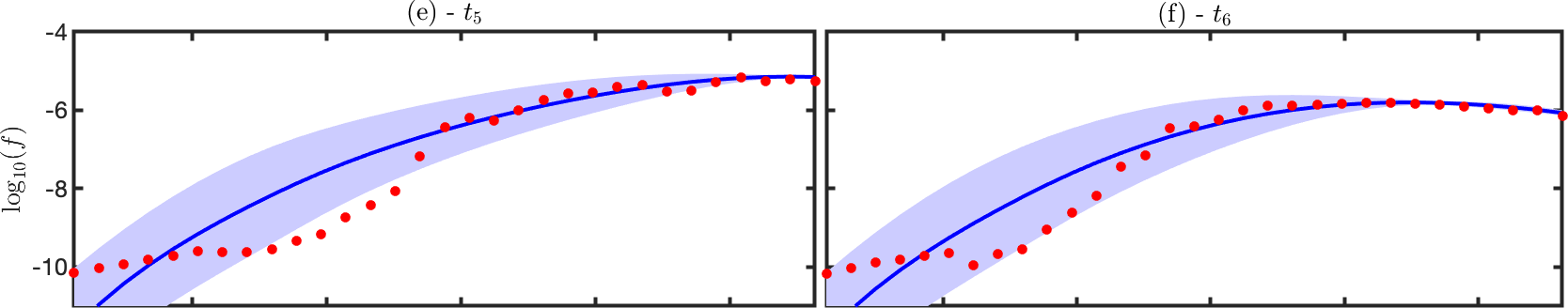} \\[\abovecaptionskip]
  \end{tabular}
  \begin{tabular}{@{}c@{}}
    \includegraphics[width=0.97\linewidth]{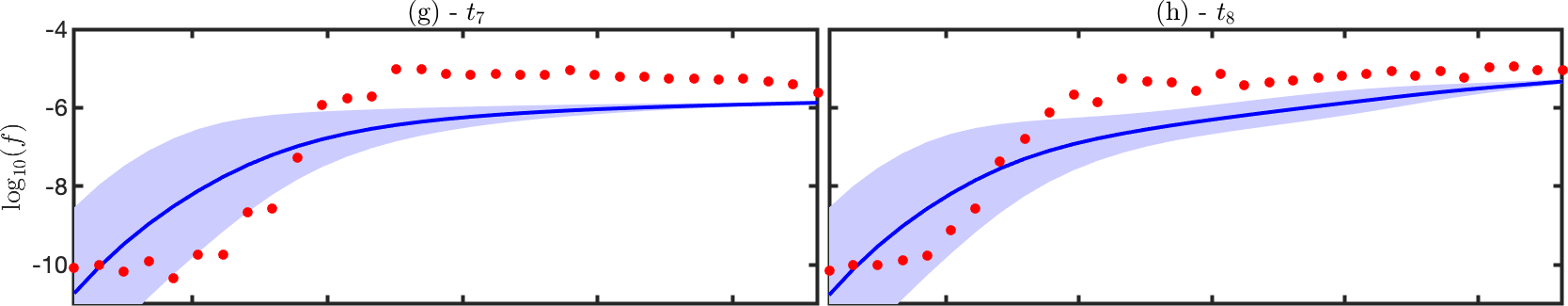} \\[\abovecaptionskip]
  \end{tabular}
  \begin{tabular}{@{}c@{}}
    \includegraphics[width=0.97\linewidth]{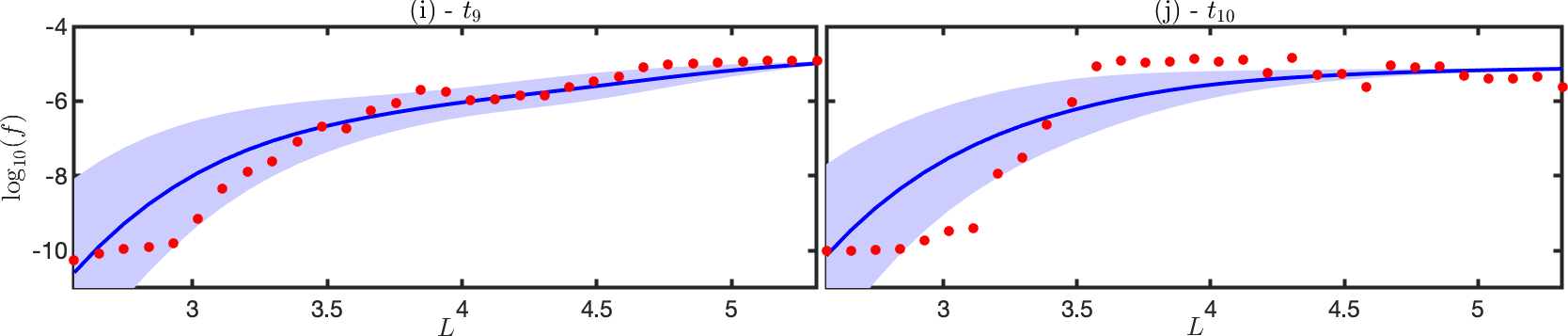} \\[\abovecaptionskip]
  \end{tabular}
  \caption{Snapshots of PSD at $\mu = 700$ MeV/G and $K=0.0019$ $\textrm{G}^{0.5}\cdot$Re for different time instances. Top panel shows the scatter plot of Van Allen Probes measurements with red lines indicating the time instances at which the probabilistic response is plotted, advancing in time from (a) to (j). Each subfigure from (a) to (j) shows the posterior-propagated PSD with mean (solid blue line), $1\sigma$ confidence intervals (blue shade) and the Van Allen measurements (red dots) corresponding to the time instance $t_1$ to $t_{10}$ in the top panel.}\label{fig:snapshots_psd}
\end{figure}

In order to visualise the probabilistic field of density in the $(L,t)$ domain, we plot the PSD at various time instances. Figure \ref{fig:snapshots_psd} shows the PSD variation at multiple time snapshots, advancing in time from Figure \ref{fig:snapshots_psd}a to \ref{fig:snapshots_psd}j. The solid line represents the mean of the posterior predictions, while the shaded area shows the $1\sigma$ confidence interval. At the beginning of the simulation near the initial condition (Figure \ref{fig:snapshots_psd}a), it can be observed that the uncertainty is low. Also there is a good agreement with respect to the Van Allen Probe  measurements (red dots), which is expected since the problem is initialised with the data. The low uncertainty is also expected since the identified posteriors have low variance. 

The solution is further time-marched and the uncertainty progressively increases (Figures \ref{fig:snapshots_psd}b-\ref{fig:snapshots_psd}e). It can be observed that the uncertainty near the upper boundary is smaller, while it is higher around the lower boundary at all time instances. This is due to the influence of the upper boundary condition, which is interpolated from the Van Allen Probe data. It is interesting to observe the influence of the injection of particles from the outer boundary. In Figure \ref{fig:snapshots_psd}f, the uncertainty around $L=4$ is observable, while after the influx of particles in Figure \ref{fig:snapshots_psd}g, the uncertainty reduces significantly. Also the Van Allen Probe data has significant noise and discontinuities in Figure \ref{fig:snapshots_psd}g, since higher density gradients diminishes the smoothness of interpolation. The uncertainty again starts to increase in Figure \ref{fig:snapshots_psd}h and a similar effect with injection of particles can be observed in Figures \ref{fig:snapshots_psd}i - \ref{fig:snapshots_psd}j. The red dots indicate the interpolated value of the Van Allen Probes data at a given time. Due to the effect of interpolation, it is not expected that the simulation outputs and the red dots would coincide perfectly. Overall the model performs well in terms of agreement with the Van Allen Probes measurements, often predicting within one standard deviation. Furthermore, the lack of agreement at lower values of $L$ is accounted by the confidence intervals. It is observed that the model discrepancy is higher when there is sudden influx of particles, which cannot be properly accounted for in a diffusive process. However, with our probabilistic approach, we are able to obtain overlap with respect to the data at most values of $L$.       

\subsubsection*{Continuous Rank Probability Score (CRPS)}
\begin{figure}
  \centering
   \includegraphics[width=1\linewidth]{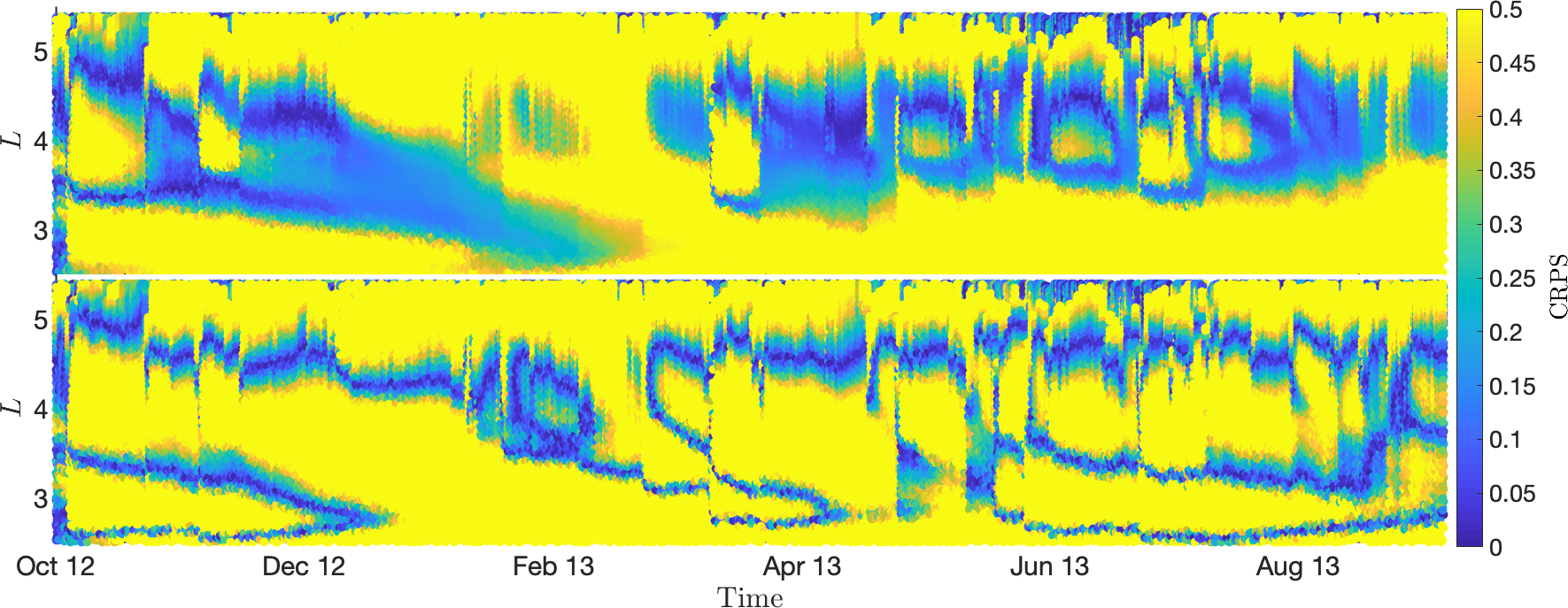} \\[\abovecaptionskip]
  \caption{Scatter plot of CRPS for the time period from October 2012 to September 2013 from Van Allen Probes. The top panel correspond to the probabilistic predictions of PSD by \eqref{eq:Diffusion_1-D} with the identified posteriors, while the bottom panel is obtained from the deterministic predictions of \eqref{eq:Diffusion_1-D} with the reference parameterization \cite{Drozdov2017}.}\label{fig:crps_compare}
\end{figure}

\begin{figure}
  \centering
   \includegraphics[width=0.4\linewidth]{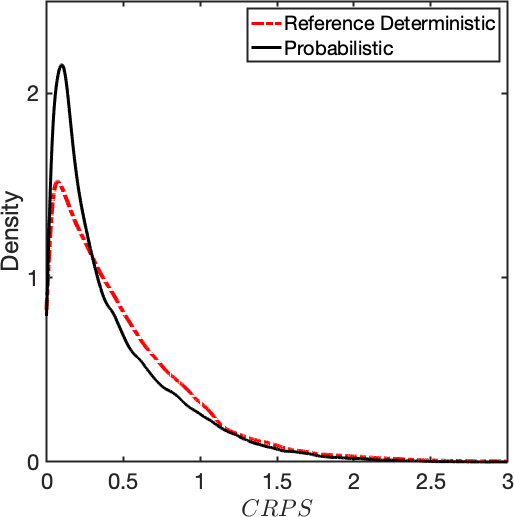} \\[\abovecaptionskip]
  \caption{Density plot of CRPS samples shown in Figure \ref{fig:crps_compare}. The black solid line correspond to the probabilistic predictions of PSD by \eqref{eq:Diffusion_1-D}, while the red dotted line correspond to predictions by the reference parameterization \cite{Drozdov2017}.}\label{fig:crps_density}
\end{figure}

In order to measure the performance of the probability forecasts, we employ the Continuous Rank Probability Score (CRPS), which is a metric to evaluate the quadratic discrepancy between the cumulative distribution function (CDF) $F$ of the forecasts and the empirical CDF of the data or observations $F^d$, which is given by:
\begin{equation}
    CRPS = \int_{-\infty}^{+\infty}\big(F(f)-F^d(f)\big)^2 df, 
    \label{eq:CRPS_formula}
\end{equation}
where $F^d$ is a Heaviside function for a scalar observation. Thus, CRPS is computed by integrating the area between the two distribution functions. If the computed area is low, the forecasts are close to the observations. Also, CRPS is a convenient measure to compare the performance of deterministic and probabilistic forecasts \cite{camporeale2019generation}. If the forecast is deterministic, CRPS reduces to a mean absolute error. Because we have a probability density of the PSD for any point of the domain $(L,t)$, we can numerically calculate the CRPS over the whole domain.
Figure \ref{fig:crps_compare} compares the CRPS obtained from the probabilistic predictions of PSD with the identified posteriors shown in Figure \ref{fig:posteriors} and the deterministic predictions with the parameterization referenced in Table \ref{tab:compare_MAP_Drozdov}. With respect to assessing the quality of the predictions, values closer to zero show that the predictor is performing better. A qualitative comparison shows that the probabilistic forecast performs better at most of the locations in the $(L,t)$ domain. Figure \ref{fig:crps_density} shows the comparison of the density of CRPS in the whole domain with the probabilistic and deterministic approaches. The mean of the CRPS is $0.429$ and $0.499$, while the 1-standard deviation is $0.4$ and $0.46$ for the probabilistic and deterministic solver respectively. From Figure \ref{fig:crps_density}, it can be seen that the probabilistic approach returns a stronger peak meaning it performs better at most sample locations in the domain. It can thus be concluded that the probabilistic treatment improves the performance of the solver. 

\subsection{MAP estimates of Phase Space Density}
\begin{figure}[h]
  \centering
   \includegraphics[width=1\linewidth]{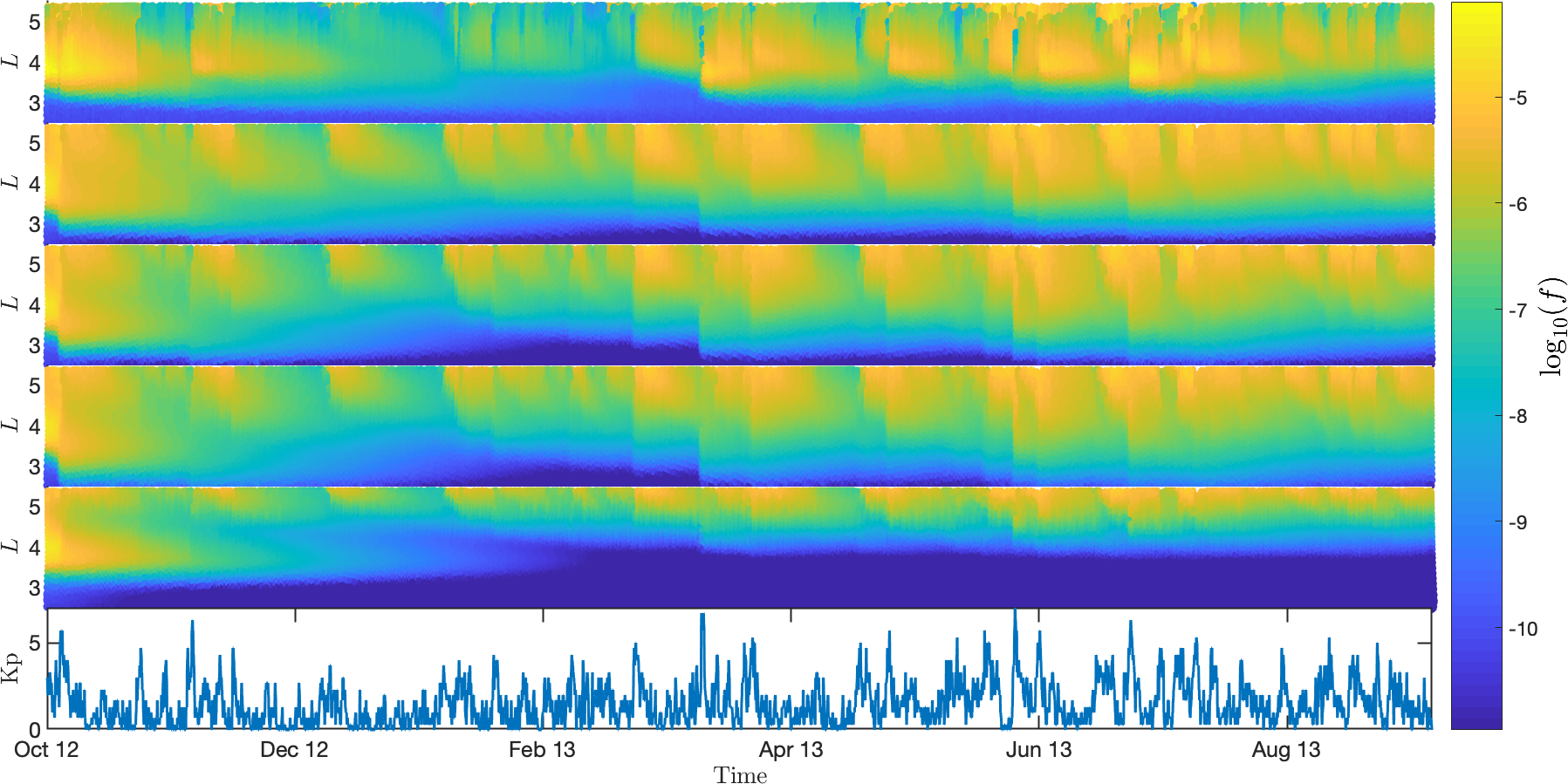} \\[\abovecaptionskip]
  \caption{Scatter plot of PSD (in logarithms of base 10) at $\mu = 700$ MeV/G and $K=0.0019$ $\textrm{G}^{0.5}\cdot$Re in $L-t$ domain for the time period from October 2012 to September 2013 obtained from Van Allen Probes measurements (top panel), model predictions by \eqref{eq:Diffusion_1-D} and parameterized by: the MAP-estimate of posteriors (second panel), reference \cite{Drozdov2017} (third panel), \eqref{eq:DLL_Ozeke} from \citeA{Ozeke2014} (fourth panel), \eqref{eq:DLL_Ali} from \citeA{Ali2016} (fifth panel), and $Kp$ index from OMNIWeb database (bottom panel).}\label{fig:psd_estimate_compare}
\end{figure}

The advantage of the proposed method clearly stays in the ability of deriving a probabilistic estimate of the PSD and the consequent ability of identifying cases of large uncertainty. However nothing precludes to use the information on the posterior distribution of the parameters in a deterministic way, by running a simulation corresponding to the MAP estimate of the parameters. In this section, we compare the PSD at $\mu = 700$ MeV/G and $K=0.0019$ $\textrm{G}^{0.5}\cdot$Re, predicted by the MAP estimate of the identified posteriors of $\mathbf{\Lambda}$ to the Van Allen Probe measurements. As a reference we again employ the 1-d model comparison provided in \citeA{Drozdov2017} (see Table \ref{tab:compare_MAP_Drozdov}). Additionally, we compare the predictions to other parameterizations existing in literature. We define $D_{LL}$ based on parameterizations obtained in \citeA{Ozeke2014} and \citeA{Ali2016}. In \citeA{Ozeke2014}, $D_{LL}$ is defined as the sum of azimuthal electric field $D_{LL}^E$ and compressional magnetic field $D_{LL}^M$, given by:
\begin{equation}
\begin{gathered}
    D_{LL}^E = L^6 2.6\cdot 10^{-8} 10^{(0.217L+0.461Kp)},\\
    D_{LL}^M = L^8 6.62\cdot 10^{-13} 10^{(-0.0327L^2+0.625L-0.0108Kp^2+0.499Kp)}.
\end{gathered}
\label{eq:DLL_Ozeke}
\end{equation}
In \citeA{Ali2016}, the drift-averaged power spectral densities are used to obtain the electric and magnetic components of the radial diffusion coefficient, and a genetic algorithm is used to derive a simple model for each component in the least squares sense, which is given by:
\begin{equation}
\begin{gathered}
    D_{LL}^E = \exp{(-16.951+0.181\cdot Kp \cdot L+1.982\cdot L)},\\
    D_{LL}^M = \exp{(-16.253+0.224\cdot Kp \cdot L +L)}.
\end{gathered}
\label{eq:DLL_Ali}
\end{equation}
Figure \ref{fig:psd_estimate_compare} shows the PSD in the form of scatter plots and compares five different realisations. It can be observed that the forward model estimations have good qualitative agreement with the Van Allen measurements. The model predictions by \eqref{eq:Diffusion_1-D} with the MAP-based and the reference parameterizations are in close agreement. This is expected since the MAP estimate of the posteriors, in particular the diffusion parameters, match closely with the reference values, which can be seen in Table \ref{tab:compare_MAP_Drozdov}. The PSD estimates from the diffusion model based on \eqref{eq:DLL_Ozeke} \cite{Ozeke2014} are also close to the MAP-based model predictions. For the diffusion model based on \eqref{eq:DLL_Ali} \cite{Ali2016}, the PSD predictions at higher values of $L$ are in qualitative agreement, however there is higher discrepancy at lower $L$ values. It has to be mentioned here that the model \eqref{eq:DLL_Ali} was calibrated in the range $3.0\leq L \leq 5.5$ and $0\leq Kp \leq 5$. Hence in this case we are performing extrapolation, which may lead to lower performance of the model.

\begin{figure}[h]
  \centering
   \includegraphics[width=1\linewidth]{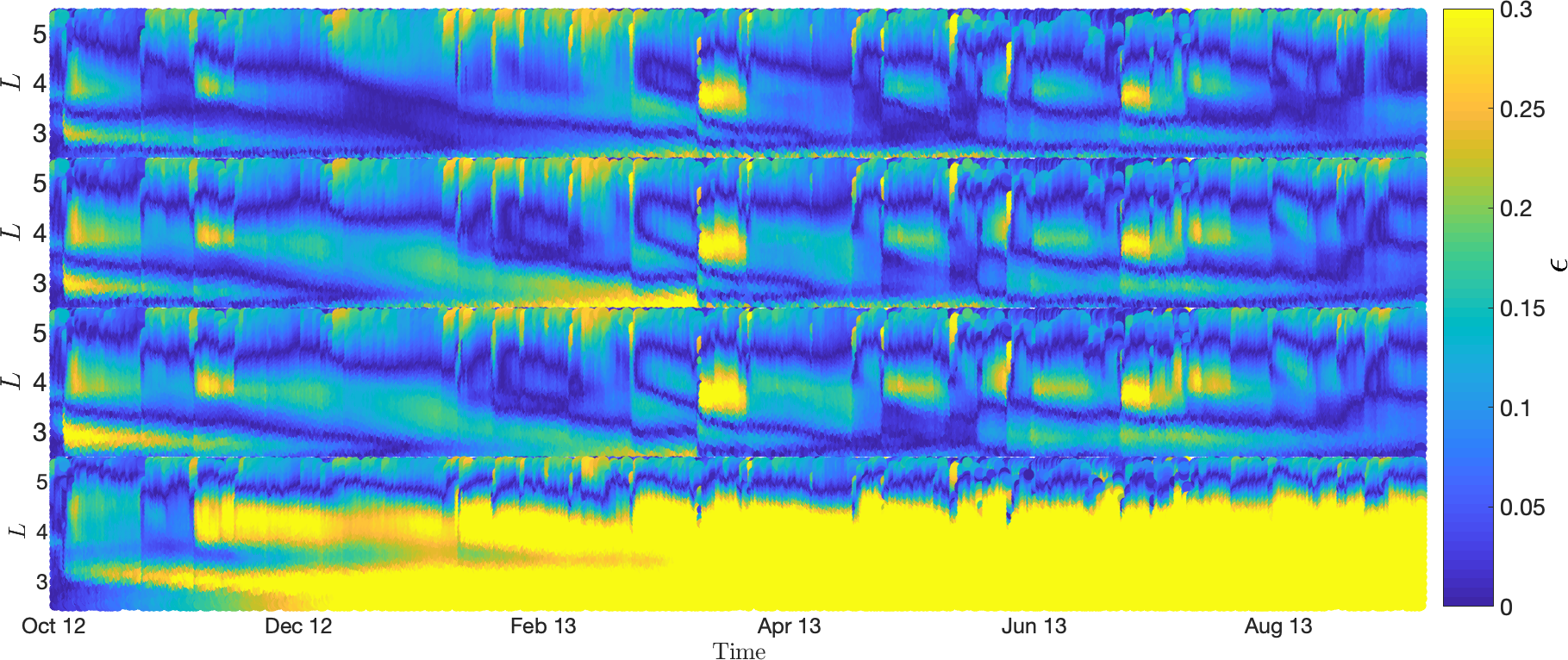} \\[\abovecaptionskip]
  \caption{Scatter plot of relative error $\epsilon$ in $L-t$ domain given by \eqref{eq:rel_err}, in estimation of PSD for the time period from October 2012 to September 2013 based on model prediction by \eqref{eq:Diffusion_1-D} and parameterized by: the MAP-estimate of posteriors (top panel), reference \cite{Drozdov2017} (second panel), \eqref{eq:DLL_Ozeke} from \citeA{Ozeke2014} (third panel), and \eqref{eq:DLL_Ali} from \citeA{Ali2016} (bottom panel).}\label{fig:rel_err_compare}
\end{figure}

We also compare the relative error given by \eqref{eq:rel_err} in estimation of PSD with scatter plots shown in Figure \ref{fig:rel_err_compare}. As expected, predictions from the MAP-based and reference parameterization match closely. The MAP-based model performs better at lower values of $L$. Similarly, the model from \citeA{Ozeke2014} also exhibits similar error behaviour. As expected, the model from \citeA{Ali2016} does not perform well at lower values of $L$. Overall, in terms of the average relative error $\epsilon_{avg}=\sum_i \epsilon_i/m$, where $m$ is the number of Van Allen Probes measurements, it is found that $\epsilon_{avg} = 0.067$, $0.074$, $0.078$ and $0.326$ with the MAP-based, reference parameterization, based on \cite{Ozeke2014}, and \cite{Ali2016} respectively. It can thus be concluded that the MAP-based model performs better than all the compared models and a small improvement is obtained if the MAP-based parameter values are employed in a deterministic setting.

\section{Conclusion}
In this paper, we demonstrated a probabilistic framework for estimating the PSD of energetic particles in the magnetosphere, following a standard quasi-linear radial diffusion equation. The probabilistic treatment is based on the Bayesian framework, where, initially, uncertainties are assumed on the parameterization of the diffusion and loss terms, which are iteratively updated using the Van Allen Probes measurements as ground truth. Finally, the informed posteriors are propagated to obtain a probabilistic estimate of particles' phase space density. It is shown that the data is able to reduce the uncertainties considerably, and the probabilistic treatment improves the quality of the predictions. This approach provides an alternate approach to understanding the 1-d radial diffusion, since a deterministic form of parameterization may not be able to represent the non-linear behaviour of this boundary-dependent problem. 

In terms of the choice of the prior, \citeA{Ali2016} obtained uncertainty bounds and attempted to define the probability distribution of the total diffusion coefficient. This could be useful to replace the uniform prior in future investigations, in order to define the domain and improve the convergence of the Bayesian model. The Bayesian estimator predicted longer time scales for the lifetime parameter, to take into account the local acceleration, which is omitted in the 1-d model. In the future, a similar approach can be extended to 3-D modeling that includes calculation of loss and local acceleration from physics. That would help better quantify the relatively unknown radial diffusion rates. Inferior radial diffusion rate is a very challenging task. In-situ measurements do not allow for inferring the m-numbers of waves or observing global distribution of waves in MLT and L. Ground measurements allow to find more global maps but observations on the ground may be very inaccurate during disturbed conditions, as ionosphere may shield the waves and they may also be guided to another location, and it may be difficult to trace back the ground observations into space. The Bayesian parameter estimation certainly has a great potential to shed light on this important problem.

\newpage



\clearpage
\newpage
\acknowledgments
The $Kp$ dataset is obtained from the Omniweb database (\url{https://omniweb.gsfc.nasa.gov}). The data generated for this research is available online at: \url{https://doi.org/10.4121/uuid:399b9a42-f07e-4b30-a4aa-86abfeeddac0}.  
\bibliography{agubiblio}

\end{document}